# Are proteins ultrametric?

Vladik A. Avetisov and Albert Kh. Bikulov

The Semenov Institute of Chemical Physics RAS, Moscow, the Russian Federation
avetisov@chph.ras.ru, bikulov1903@rambler.ru

The question posed in the title of this article has been raised by Hans Frauenfelder over 20 years ago [1]. When studying the ligand-rebinding kinetics of myoglobin, Frauenfelder discovered that below the room temperature, at which the protein dynamics limit the rebinding rate, the reaction kinetics obeys a power law or a stretched exponent [2]. In order to explain this fact, he supposed that the protein energy landscape (EL) has a great number of local minima corresponding to conformational substates (CSs) with nearly the same energies. With respect to the transition rates between CSs, the local minima were assumed to be clustered into hierarchically embedded basins of minima: the smaller basins separated by smaller activation barriers are pooled into larger basins that are separated by higher barriers. In other words, Frauenfelder suggested that the protein CSs and the protein EL clearly do exhibit some taxonomic order. Since hierarchical taxonomy can be described using non-Archimedean (ultrametric) distances, i.e. distances satisfying the strong triangle inequality, an ultrametric space is introduced to describe the protein CSs. Hence, the protein dynamics are associated with an ultrametric random process. The term "protein ultrametricity" should be understood in the same sense.

Frauenfelder's hypothesis, which has attracted a great deal of interest (see, for example [3,4]), is regarded as one of the most profound ideas put forward to explain the nature of protein attributes that has been proposed in the last decades. To date, however, no theoretical validation of protein ultrametricity has been found that would be accepted by the whole scientific community. In earlier theoretical works [5,6], some models were proposed to describe a random walk over an ultrametric space (ultrametric diffusion), but these models were confronted with difficulties in applications to the ligand-rebinding kinetics of myoglobin. More consistent ultrametric approaches were developed in [7,8], where *p*-adic numbers and *p*-adic pseudo-differential equations were used to describe ultrametric diffusion. It was shown in [7,8] that the specific features of the ligand-rebinding kinetics of myoglobin can be described on the basis of Frauenfelder's hypothesis on the protein ultrametricity.

In this article we discuss the spectral diffusion in deeply frozen proteins in the same context. In particular, we show that unfamiliar features of spectral diffusion in proteins also can be described on the basis of ultrametricity of protein dynamics.

Spectral diffusion in proteins has been studied for over two decades and has offered a unique experimental background to a thorough insight into the subtle interplay between order and disorder in proteins (for earlier data see [9]; for recent reviews see [10-12]). Spectral diffusion in proteins is studied via measurement of the optical absorption spectrum of a chromophore marker, implanted, or naturally occurring in a heme protein. The marker absorption frequency is highly sensitive to the arrangement of its local surroundings: displacements of the marker's neighbors of about $\sim 10^{-2}$ Å may result in frequency shifts of about ~0.1 GHz (for the estimations see, for example, [13]). At low temperatures, the typical time-scales between such frequency shifts are long enough to allow the observation of the shifts as a random process over time, yet the absorption spectrum of a macroscopic sample is inhomogeneously broadened. The typical width of an adsorption band is $\sim 10^3$ GHz at 4.2 K; therefore, the random frequency shifts of individual markers are buried in the sample absorption spectrum. In order to observe the random process, one can use markers that undergo an irreversible photochemical transition in the case of impulse laser pumping near the maximum of the absorption spectrum. Then a part of the markers is "blinded" and a narrow spectral hole is burned in the absorption spectrum. Because of the fact that the frequency of the optical transition of blinded markers is being randomly changed due to rearrangements of the marker's neighbors, the burnt spectral hole broadens with time, producing a spectral diffusion. The features of this random process are established by monitoring the changes of the burnt hole (its width, $\sigma$) as a function of the aging time, $t_{ag}$, and the waiting time, $t_w$. The aging time, $t_{ag}$, is the interval between the moment when the deeply frozen sample is prepared, e.g. the time-point when both the sample temperature and the absorption spectrum are thought to be stabilized, and the time-point when the spectral hole is burnt. The waiting time $t_w$ is counted from the time-point when the spectral hole is burnt, and it is the current time for the spectral diffusion.

Spectral diffusion has been studied for various heme proteins, both with a native and a denatured structure, various organic environments of proteins in samples, and for various temperatures and temperature cycles. The constitutive properties of the low-temperature spectral diffusion in native proteins are universal and were, for the most part, discovered by J. Friedrich's group. They are as follows (see, for example, [9]). ***Spectral diffusion kernel*** – In nearly the entire time window of the spectral diffusion monitoring from $t_w \sim 10\,\text{min}$ up to $t_w \sim 10^4\,\text{min}$, the spectral hole shape is satisfactorily approximated by the Gaussian. Thus, the spectral diffusion in proteins can be regarded as a Gaussian random process propagating along the frequency straight line. ***Spectral hole broadening*** – In proteins with a native structure, the increase of the Gaussian width of the spectral hole with the waiting time, $\sigma(t_w)$, is well approximated by the power-law function $\sigma(t_w) \sim t_w^a$ (**Fig.1**, at the top). It is worth noting that the exponent $a$ is not equal to 0.5, in contrast to the familiar diffusion; its value (estimated from a power-law fit) falls into the narrow interval $a = 0.25 \div 0.30$, depending on the samples selected. Thus, in the case of native proteins, spectral diffusion differs dramatically from both familiar diffusion and spectral diffusion in organic glasses, where it is broadened logarithmically with waiting time (for details see [9,14]). It should be emphasized that the spectral diffusion in denatured proteins subjected to rather long aging ($t_{ag} \approx 10^2\,\text{h}$ and more) is broadening as $\sim \ln t_w$ [11]; i.e. at low temperatures, denatured proteins are similar to organic glasses, although native proteins are not. ***Spectral diffusion aging*** – In native proteins, the broadening of the spectral hole depends also on the aging time. The greater the value of $t_{ag}$ the smaller the value of $\sigma(t_w)$ at the same $t_w$. For aging times from $t_{ag} \sim 1\,\text{h}$ up to $t_{ag} \sim 10^2\,\text{h}$, this decrease at the waiting time-point $t_w = 10^4\,\text{min}$ ($\approx 200\,\text{h}$) is fit satisfactorily by the power function $\sigma(t_{ag}, t_w = 10^4\,\text{min}) \sim t_{ag}^{-b}$, where $b = 0.07 \pm 0.01$ (**Fig.1**, at the bottom). It is important to notice that the aging exponent $b$ proves to be also nearly constant in the case of native proteins.

Several alternative approaches based, to some extend, on similarity between proteins and glassy systems with quenched disorder, have been proposed in order to describe random processes with characteristics of spectral diffusion in deeply frozen proteins (see [10-12]). These approaches catch the fact that broad distributions of the relaxation times are typical of proteins and glasses and, on the other hand, such specificity can result in anomalously slow diffusion and aging (see, for example, [15,16]). For these reasons, the spectral diffusion in protein itself, i.e. the random shifts of the marker absorption frequency, has been described in terms of the trap models, the two-level-system models, and the random walk along the random pass (see [10-12]). However, such approaches find difficulty in decoding information about the protein CSs and the protein EL, since the protein dynamics appear out of consideration. In fact, as we show below, an accurate and surprisingly simple description of spectral diffusion in native proteins can be constructed directly on the basis of ultrametricity of protein dynamics. In our approach, the features of spectral diffusion in proteins are resulting from the protein dynamics, in contrast to the approaches mentioned above. It is important to note also that the description of protein dynamics in our approach is similar to that used earlier for describing the ligand-rebinding kinetics of myoglobin [7]. Thus, we show that Frauenfelder's idea offers a universal background for description of protein dynamics in a very wide range of protein motion scales.

The ultrametric description of the protein dynamics can be outlined as follows (for details see [7,8]). Let us introduce an ultrametric ($p$-adic) tree with a branching index $(p+1)$ at the tree nodes, $p$ is a prime integer (Fig.2). With the Frauenfelder's hypothesis in mind, we associate the lowest nodes of the tree with the protein CSs (local minima of the protein EL), and the other nodes (branching points) are associated with the saddle points of EL lying at the transition paths between CSs. Such a tree can be regarded as a graph of ultrametric distances between the protein CSs and, on the other hand, as a hierarchical "skeleton" of the protein EL representing hierarchically embedded basins of CSs. A transition rate between two CSs is determined by the level $\gamma$, at which a vertex of the minimal subtree with given CSs is located.

Let us model the protein dynamics on the CSs by an ultrametric diffusion, i.e. random walk on the bottom nodes of the ultrametric tree. In order to write the *p*-adic equation of ultrametric diffusion, one can begin with the master equation of a form [5]:

$$\frac{d\mathbf{F}(t)}{dt} = \mathbf{W}\mathbf{F}(t) \qquad (1)$$

where $\mathbf{F}(t) = \{f_1, \ldots, f_N\}$ is interpreted as the protein state vector defined on the protein CSs, i.e. the bottom nodes of the ultrametric tree, $f_i(t)$ is the probability to be in the *i*-th CS at a time *t*, and the transition matrix $\mathbf{W}$ has a block structure similar to that of the Parisi matrix (**Fig. 3**). Such a block structure of $\mathbf{W}$ relates to ultrametricity of the protein CSs with respect to the transitions between them. The elements of the transition matrix $\mathbf{W}$ can be parameterized by a set of *p*-adic numbers $x \in Q_p$ [17,18], whose norm satisfy the strong triangular inequality. Then, the protein CSs can be indexed by *p*-adic numbers, the ultrametric distances between CSs can be determined by the *p*-adic norm $|x-y|_p$, $x, y \in Q_p$, and the transition rates given by the elements of the matrix $\mathbf{W}$ can be specified by a well-defined function of ultrametric distance, $w_{xy} = w(|x-y|_p)$. In the continuum limit, i.e. continuation of the ultrametric tree in both "up" and "down" directions, one can transform the master equation (1) to a *p*-adic equation of ultrametric diffusion [7,17]):

$$\frac{\partial f(x,t)}{\partial t} = \int_{Q_p} |x-y|_p^{-(\alpha+1)} [f(y,t) - f(x,t)] d_p y \qquad (2)$$

Here, the real-valued function $f(x,t)$, $x \in Q_p$, $t \in R_+$ is the distribution of the probability density over the field of *p*-adic numbers $Q_p$ at time $t$, and $d_p y$ is the Haar measure of integration over $Q_p$. In the problem under the question the field $Q_p$ describes an ultrametric space of the protein CSs, and the kernel of the integral operator in the right hand side of equation (2), $|x-y|_p^{-(\alpha+1)} = p^{-(\alpha+1)\gamma}$, describes the protein EL in terms of the transition rates between the CSs. The parameter $\alpha$ in equation (2) scales the hierarchy of the transition rates. In the general case, the kernels of ultrametric diffusion can be chosen of various form [8,19], and in our case it decreases exponentially with transition crossing a higher level $\gamma$ in the ultrametric tree. As it has been shown earlier [7,8], such a dependence of the transition rates on the ultrametric distance is appropriate for description of protein dynamics in the case of the ligand-rebinding kinetics of myoglobin.

With the protein dynamic equation (2) in hands, we are in position to discuss spectral diffusion in deeply frozen proteins. First of all, we elucidate qualitatively the key question, namely, how random changes in the marker absorption frequency are interrelated with random changes in the protein CSs. Or, in other words, how the spectral diffusion in protein is coupled with the protein dynamics. In order to understand this relation, it is sufficient to compare the number of frequency-distinguishable configurations of the marker neighbors and the number of the protein CSs. An estimate of the first of these numbers can be obtained using the ratio between the sample absorption band ($\sim 10^3$ GHz) and the absorption line-width of a marker ($\sim 0.1$ GHz). This gives $10^4 \approx 2^{13}$ lines in the sample absorption spectrum; i.e. the marker neighbors constitute a subsystem with about 10 degrees of freedom, each with a few frequency-distinguishable states. In contrast, a protein CS has hundreds of degrees of freedom, and so the protein CS-space is "astronomically" large: the number of CSs can be as large as $2^{100}$. Although these estimates are of a symbolic nature, when comparing $2^{13}$ and $2^{100}$, we can certainly make a conclusion that "almost all" of the transitions between CSs do not result in changes of the marker absorption frequency. Thus, the spectral diffusion is due to rare random events occurring in the midst of transitions between the protein CSs. Such rare events can be associated with hitting very particular CSs. We shall call such CSs "zero-points" of the protein dynamic trajectory, and the series of time-points, when the trajectory hits zero points, "zero-point clouds". Thus, the spectral diffusion in proteins can be regarded as a Gaussian random process, whose development in time is given by zero-point clouds of the protein dynamic trajectory. We stress that spectral diffusion itself is propagating along the frequency straight line, while the protein evolves in the CS-space, which is an ultrametric "straight line" in our approach. As zero points are very peculiar CSs, we shell model them by a small ultrametric ball $Z_p \left( x \in Z_p, |x|_p \leq 1 \right)$ inside an infinite space $Q_p$.

Let us now specify the spectral diffusion as a frequency random walk as follows:

$$v(\tau) = \overline{v} + \sum_{i=1}^{n(\tau)} \Delta v_i \qquad (3)$$

where $\overline{v}$ is the mean frequency averaged over the Gaussian spectral hole, and $\Delta v_i$, $i = 1,\ldots,n(\tau)$, are independent frequency increments. For the sake of simplicity, we consider $\Delta v_i$ as a "telegraph noise" taking values $\Delta v_i = \pm \Delta$ with equal probability of 0.5. The random intender $n(\tau)$ is the number of times the protein dynamic trajectory hits the zero-point CSs, i.e. the ultrametric ball $Z_p$, during the time interval $\tau = [t_{ag}, t_{ag} + t_w]$.

Let $P(n(\tau))$ be the probability distribution for $n(\tau)$. Then the mean-square deviation $\sigma^2(t_{ag}, t_w)$ of process (3) is written as:

$$\sigma^2(t_{ag}, t_w) = \langle\langle \left(\sum_{i=1}^{n(\tau)} \Delta v_i\right)^2 \rangle_{\Delta v}\rangle_{n(\tau)} = \Delta^2 \langle n(\tau) \rangle_{n(\tau)} \qquad (4)$$

Thus, all we need for the description of spectral diffusion is to find the average hitting $Z_p$, i.e. the value $\langle n(\tau) \rangle_{n(\tau)}$, via ultrametric diffusion during the time interval from $t = t_{ag}$ until $t = t_{ag} + t_w$. The solution of this problem needs detailed analysis of zero-point clouds of ultrametric diffusion [20], and the summary of the analysis is as follows. Using equation (2), it is easy to write the equation for the probability $S(t) = \int_{Z_p} f(x,t) d_p x$ to find a protein at the CSs belonging to the ultrametric ball $Z_p$:

$$\frac{\partial S(t)}{\partial t} = -B(\alpha)S(t) + G(t) \qquad (5)$$

where $B(\alpha) = (1 - p^{-1})/(1 - p^{-(\alpha+1)})$, and $G(t) = \frac{1 - p^{-\alpha}}{1 - p^{-(\alpha+1)}} \int_{|y|_p > 1} |y|_p^{-(\alpha+1)} f(y,t) d_p y$.

Since $G(t)$ is the mean rate of transitions into $Z_p$ at time $t$ averaged over an ensemble, one can write the average number of hitting $Z_p$ during the time interval $\tau = [t_{ag}, t_{ag} + t_w]$ in the form:

$$\overline{n}(t_{ag}, t_w) = \int_{t_{ag}}^{t_{ag}+t_w} G(t)dt = B(\alpha) \int_{t_{ag}}^{t_{ag}+t_w} S(t)dt + [S(t_{ag} + t_w) - S(t_{ag})] \qquad (6)$$

The probability $S(t)$ can be found from the solution of the Cauchy problem for equation (2) and the initial condition

$$f(x,0) = \begin{cases} 1, & \text{if } |x|_p \leq 1 \\ 0, & \text{if } |x|_p > 1 \end{cases}.$$

It has the following form [7,8]:

$$S(t) = (1-p^{-1})\sum_{\gamma=0}^{\infty} p^{-\gamma} \exp(-p^{\alpha\gamma}t) \tag{7}$$

Then combining (4), (6), and (7), we can find out an exact analytic expression for the mean-square deviation $\sigma^2(t_{ag}, t_w)$. For the values of $\alpha \approx 2$ and $\Delta \approx 0.3$ GHz, this expression describes very well both the broadening and the aging of spectral diffusion in deeply frozen proteins (**Fig.4a,b**). Note, that the spectral diffusion aging in proteins is caused in our description exactly by the protein dynamic specificity that J.-P. Bouchaud called "aging on Parisi's tree" [21], and earlier H. Frauenfelder called "the protein nonergodicity" [1].

The power laws of the spectral diffusion broadening and aging relying on the fitting of experimental data (see **Fig.1**) can be obtained using reasonable approximation of the solution (7) by the power function $S(t) \sim t^{-1/\alpha}$ that was found earlier in [7,8]. Again, substituting $S(t) \sim t^{-1/\alpha}$ to equation (6), and neglecting the fast-decaying terms $\sim t^{-1/\alpha}$, one can rewrite the mean-square deviation $\sigma^2(t_{ag}, t_w)$ in the form

$$\sigma^2(t_{ag}, t_w) \sim \left[ (t_{ag} + t_w)^{\frac{\alpha-1}{\alpha}} - (t_{ag})^{\frac{\alpha-1}{\alpha}} \right] \tag{8}$$

It is an important result, as obtained in [18], that the first passage times, $\tau_f$, of ultrametric diffusion is given by the distribution $\Omega(\tau_f) \sim \tau_f^{-(2\alpha-1)/\alpha}$. As long as the probability $S(t)$ remains high enough, the statistics of hitting $Z_p$ changes with aging time $t_{ag}$, for the most part, due to the decreasing of $S(t_{ag})$, i.e. $\Omega(t_{ag}, \tau_f) \sim S(t_{ag})\tau_f^{-(2\alpha-1)/\alpha}$. This condition is satisfied if $t_{ag} < t_w$, i.e. it is just the case of the experiment (see Fig. 1, the bottom). When rescaling the distribution $\Omega(\tau_f)$ with respect to the aging time and then redefining the distribution $P(n(\tau))$, the master-curve relation can be obtained from (8):

$$\sigma(t_{ag}, t_w) \sim \left( t_{ag}^{-\frac{1}{(2\alpha-1)}} t_w \right)^{\frac{\alpha-1}{2\alpha}} \qquad (9)$$

This relation coincides exactly with the power-law fittings of experimental data presented in Ref. [9-12]. We should emphasize that the only parameter $\alpha$, which specifies the time scales of protein dynamics, determines both the exponent *a* and the exponent *b* of the power-laws of spectral diffusion broadening and spectral diffusion aging. In particular, *a* = 0.27 and *b* = 0.08 are obtained from equation (9) at $\alpha = 2.2$, and the same exponents are obtained from the exact solution $\sigma(t_{ag}, t_w)$ given by (4), (6), and (7) (see Fig.4a,b).

Thus, we consider the above results as one more convincing argument in favor of protein ultrametricity. As we have demonstrated above, Frauefelder's hypothesis offers a simple and relevant description of protein dynamics on a large range of protein motion scales. It is interesting to note that, since the hierarchy of transition rates given by the *p*-adic equation (2) describes the dynamics of native proteins surprisingly accurately, the protein CS-basins and the protein EL should exhibit the hierarchical self-similarity. This theoretical observation may be of great importance for a deeper understanding of the nature of protein ordering.

**Acknowledgments.** We express our gratitude to V. S. Vladimirov, I. V. Volovich, S. V. Kosyrev, A. P. Zubarev, and S. K. Nechaev for fruitful discussions. The authors thank K. A. Bashevoy and V. A. Ivanov for assistance in computer simulations. This research was partially supported by the RFBR (grants No: 05-03-32563a, 07-02-00612a) and the Program OCHNM RAS (1-OCH/06-07).

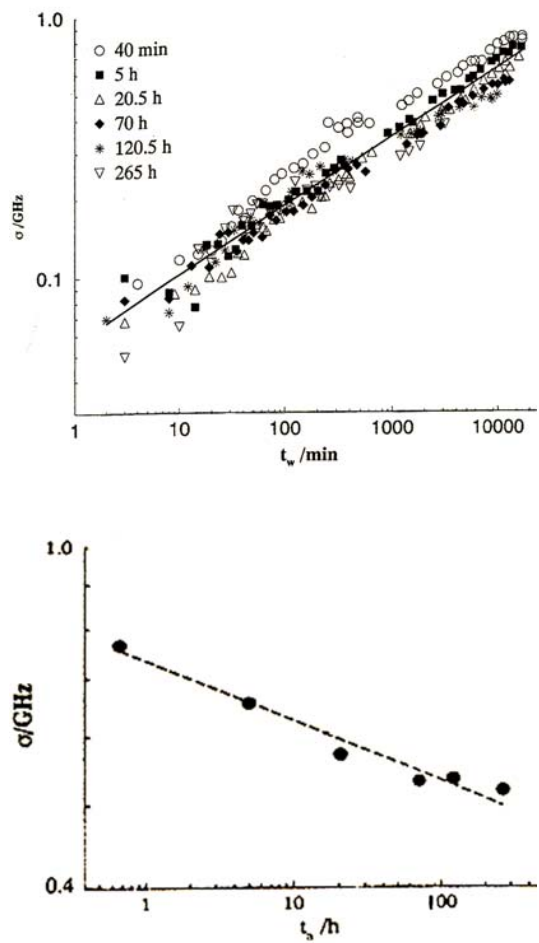

**Fig.1.** The top panel: Spectral diffusion broadening $\sigma(t_w)$ in horseradish peroxidase at 4.2 K. The solid line represents a power-law fit with an exponent $0.27 \pm 0.02$. The bottom panel: Spectral diffusion aging behavior specified at the waiting time-point $t_w = 10^4$ min for various aging times indicated by different symbols and. The dashed line represents a power-law aging fit with an exponent $-0.07 \pm 0.01$; (the data and the fits are reproduced from [9])

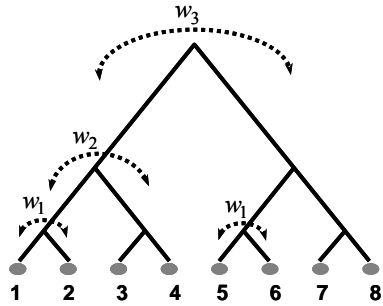

**Fig.2.** 2-adic ultrametric tree. Ultrametric diffusion is a random transitions between the lowest nodes of the tree. The rates of transitions ($w_1$, $w_2$, $w_3$) are specified by a function of ultrametric distance. As measured from node 1, node 2 has an ultrametric distance $2^1$, nodes 2 and 3 have an ultrametric distance of $2^2$, and nodes 5-8 have an ultrametric distance $2^3$.

$$\mathbf{W} = \begin{pmatrix} w_0 & w_1 & w_2 & w_2 & w_3 & w_3 & w_3 & w_3 \\ w_1 & w_0 & w_2 & w_2 & w_3 & w_3 & w_3 & w_3 \\ w_2 & w_2 & w_0 & w_1 & w_3 & w_3 & w_3 & w_3 \\ w_2 & w_2 & w_1 & w_0 & w_3 & w_3 & w_3 & w_3 \\ w_3 & w_3 & w_3 & w_3 & w_0 & w_1 & w_2 & w_2 \\ w_3 & w_3 & w_3 & w_3 & w_1 & w_0 & w_2 & w_2 \\ w_3 & w_3 & w_3 & w_3 & w_2 & w_2 & w_0 & w_1 \\ w_3 & w_3 & w_3 & w_3 & w_2 & w_2 & w_1 & w_0 \end{pmatrix}$$

**Fig.3.** Transition matrix W with a block structure similar to that of the Parisi matrix. The elements $w_1$, $w_2$, and $w_3$ are the transition rates on the ultrametric distances $2^1$, $2^2$, and $2^3$, respectively.

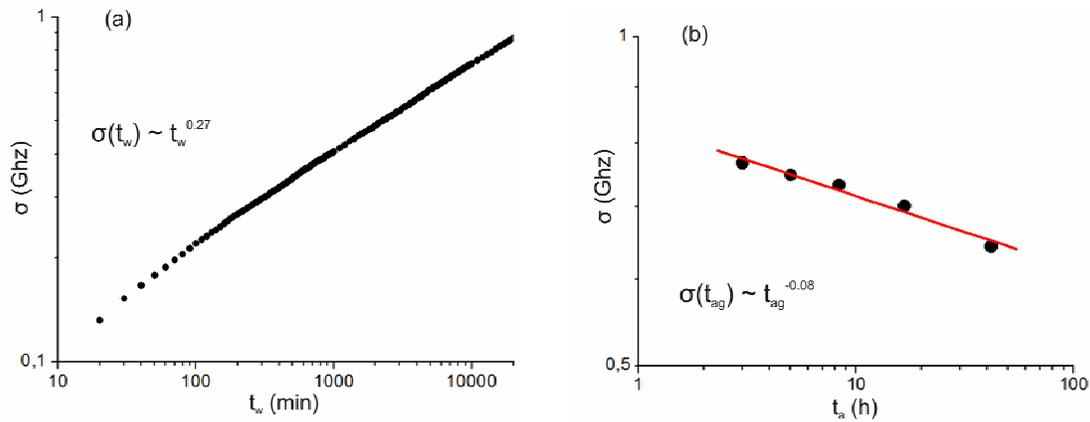

**Fig.4.** Log-log representation of the mean-square deviation $\sigma^2(t_{ag}, t_w) \sim t_{ag}^{-0.08} t_w^{0.27}$ obtained from the equations (4), (6), and (7) at $\alpha = 2.2$ and $\Delta \approx 0.3$ GHz (see text for explanations): (a) - spectral diffusion broadening; (b) spectral diffusion aging at the waiting time-point $t_w = 10^4$ min and aging from a few hours up to a hundred hours. Note, that the axis scales coincide with the real data (see Fig.1).